\documentclass[11pt,twoside]{article}
\usepackage{vdbbook}
\resetcounters
\begin{document}
\allowtitlefootnote
\title{Far infra-red emission lines in high redshift quasars}
\author{Simona Gallerani$^1$
\affil{$^1$Scuola Normale Superiore di Pisa, Piazza dei Cavalieri 7, 56126 Pisa, Italy\\
}
}
\begin{abstract}
We present Plateau de Bure Interferometer (PdBI) observations of far infra-red emission lines in BRI 0952-0115 (B0952), a lensed QSO at z=4.4 powered by a super-massive black hole ($M_{\rm BH}=2\times 10^9 M_{\odot}$). In this source, the resolved map of the [CII] emission at 158~$\mu$m allows us to reveal the presence of a companion galaxy, located at $\sim 10$~kpc from the QSO, undetected in optical observations. From the CO(5-4) emission line properties 
%(luminosity and width) are consistent with a molecular hydrogen mass $M_{\rm H_2}\sim 2\times 10^9 M_{\odot}$, and a dynamical mass $M_{\rm dyn}< 2.4\times 10^{10} M_{\odot}$. We therefore 
we infer a stellar mass $M_*<2.2 \times 10^{10} M_{\odot}$, which is significantly smaller than the one found in local galaxies hosting black holes with similar masses ($M_*\sim 10^{12} M_{\odot}$). The detection of the [NII] emission at 205~$\mu m$ suggests that the metallicity in B0952 is consistent with solar, implying that the chemical evolution has progressed very rapidly in this system. We also present PdBI observations of the [CII] emission line in SDSSJ1148+5251 (S1148), one of the most distant QSO known, at z=6.4. We detect broad wings in the [CII] emission line, indicative of gas which is outflowing from the host galaxy. In particular, the extent of the wings, and the size of the [CII] emitting region associated to them, are indicative of a QSO-driven massive outflow with the highest outflow rate ever found ($\rm \dot{M}> 3500~M_{\odot}~yr^{-1}$).
\end{abstract}
\section{Introduction}
Far infra-red emission lines represent exquisite tools for constraining the properties of the ISM in star-forming galaxies and for studying several processes related to galaxy formation (e.g. galaxy merging and SN/QSO feedback). We present PdBI observations of far infra-red emission lines in two sources, namely B0952 and S1148. B0952 is a QSO at z=4.4 (McMahon et al. 1992) lensed by a foreground galaxy at z=0.6, as revealed by the double optical images detected with NICMOS-HST (Leh\'ar et al. 2000) and by the FORS1-VLT spectrum (Eigenbrod et al. 2007). Strong [CII] emission has been discovered towards this object using APEX (Maiolino et al. 2009). In this work, we report PdBI observations of B0952 of the [CII]~158 $\mu$m, [NII]~205 $\mu$m, and CO(5-4) emission lines. Further information on the [CII] observations can be found in Gallerani et al. (2012) (G12), while [NII] and CO(5-4) results will be presented in a forthcoming paper (Gallerani et al. in preparation). S1148 at z=6.4 is the first source in which the [CII] emission has been detected (Maiolino et al. 2005) and resolved (Walter et al. 2009). We present follow-up observations of the [CII] line obtained with the PdBI, as described in full details in Maiolino et al. (2012).  
\label{intro}
\section{[CII], CO(5-4), [NII] emission in B0952}
\begin{figure}
\centering
\includegraphics[scale=0.22]{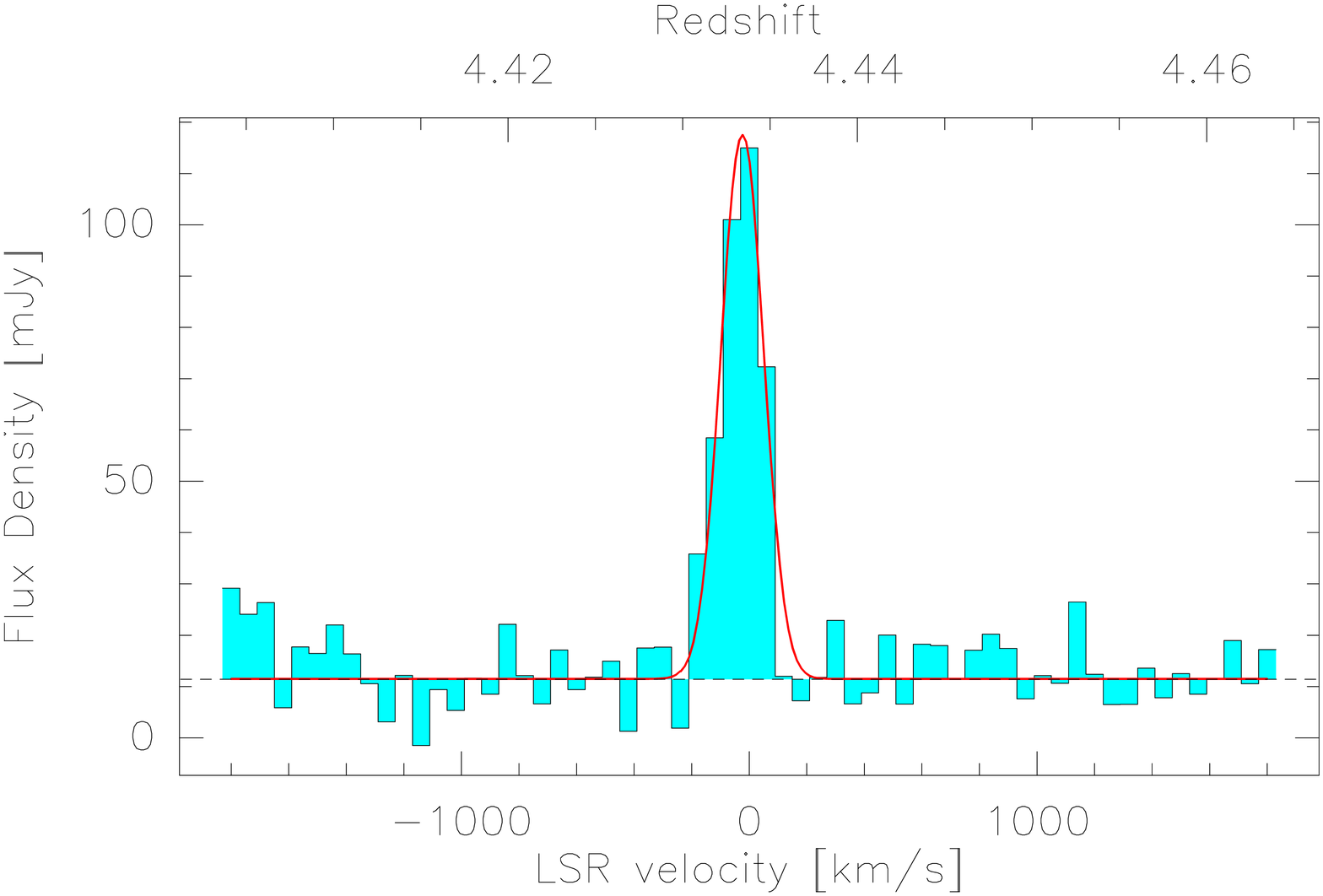} 
\includegraphics[scale=0.35]{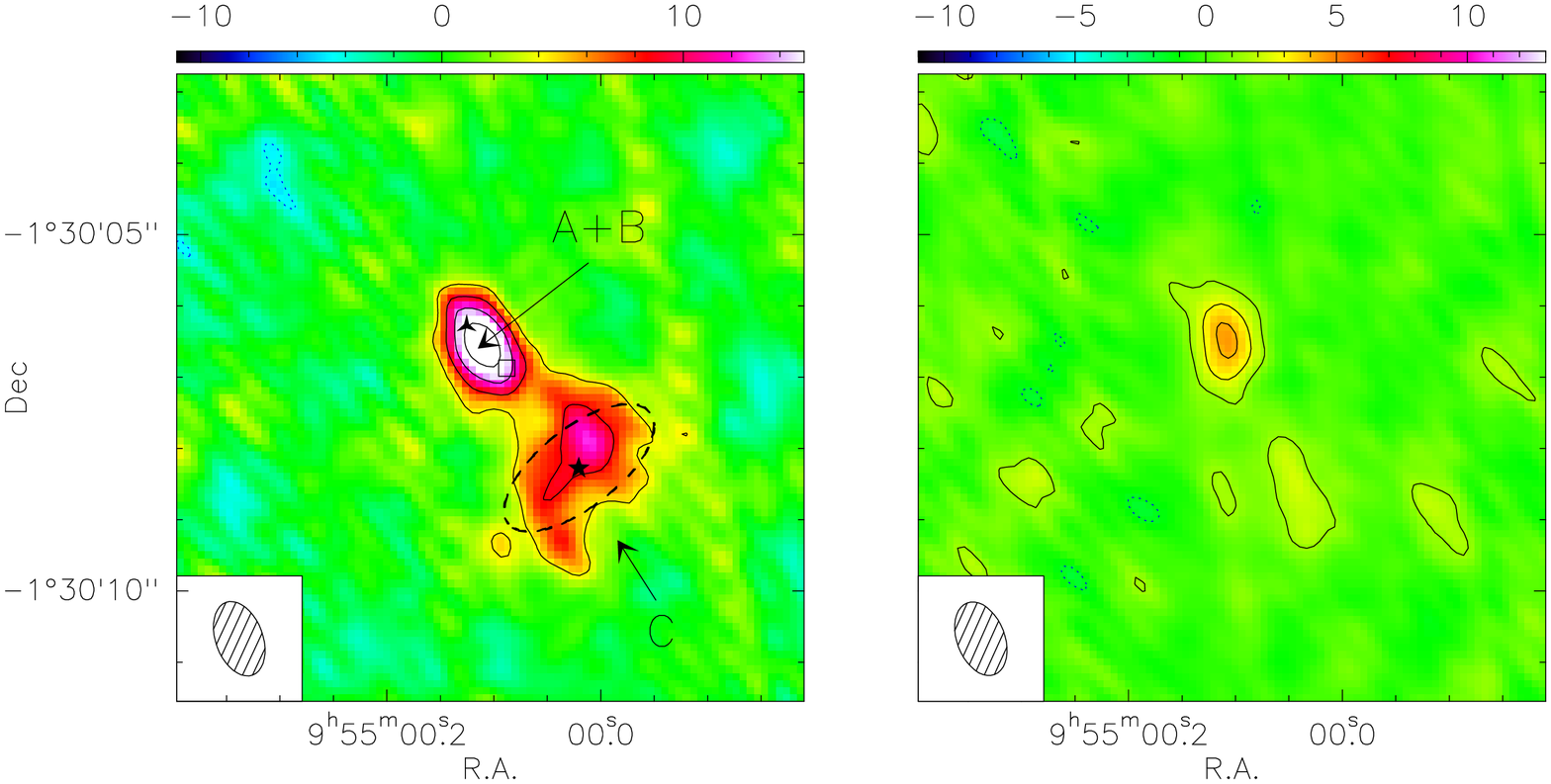}
\caption{Left panel: [CII] spectrum of B0952 obtained with the PdBI by integrating over a region which follows the 3$\sigma$ contour on the core map. The spectrum is shown on top of a continuum of $F_{cont}=11.4\pm 1.4$~mJy (see right panel). Middle panel: map of the [CII] line emission region integrated over the velocity range $\rm -210 < v < 90~[km~s^{-1}]$. Contour levels are shown in steps of 2.5$\sigma$, where 1$\sigma=0.5$ Jy~km~s$^{-1}$~beam$^{-1}$. The synthesized beam of $1.08''\times 0.66''$ is shown in the bottom-left insets. Right panel: map of the continuum emission obtained from the line-free channels of the 3.6~GHz wide spectrum. Contour levels are shown in steps of 2.5$\sigma$, where 1$\sigma=0.5$ mJy~beam$^{-1}$.} 
\label{figcii}
\end{figure}

The results of our PdBI observations are shown Fig. \ref{figcii}. Our data reveal that B0952 is characterized by a complex structure: besides the [CII] blended image of the double lensed images of the QSO nucleus (component A+B in Fig. \ref{figcii}, see also Fig. 3 in G12) we detect a second extended ($\sim 12$~kpc) region (component C) located $\sim$~10 kpc from the QSO. The C component has not been detected with HST, possibly because of dust obscuration, and it is likely a companion galaxy in the phase of merging with the QSO host. 

\begin{figure}
\centering
\includegraphics[scale=0.25]{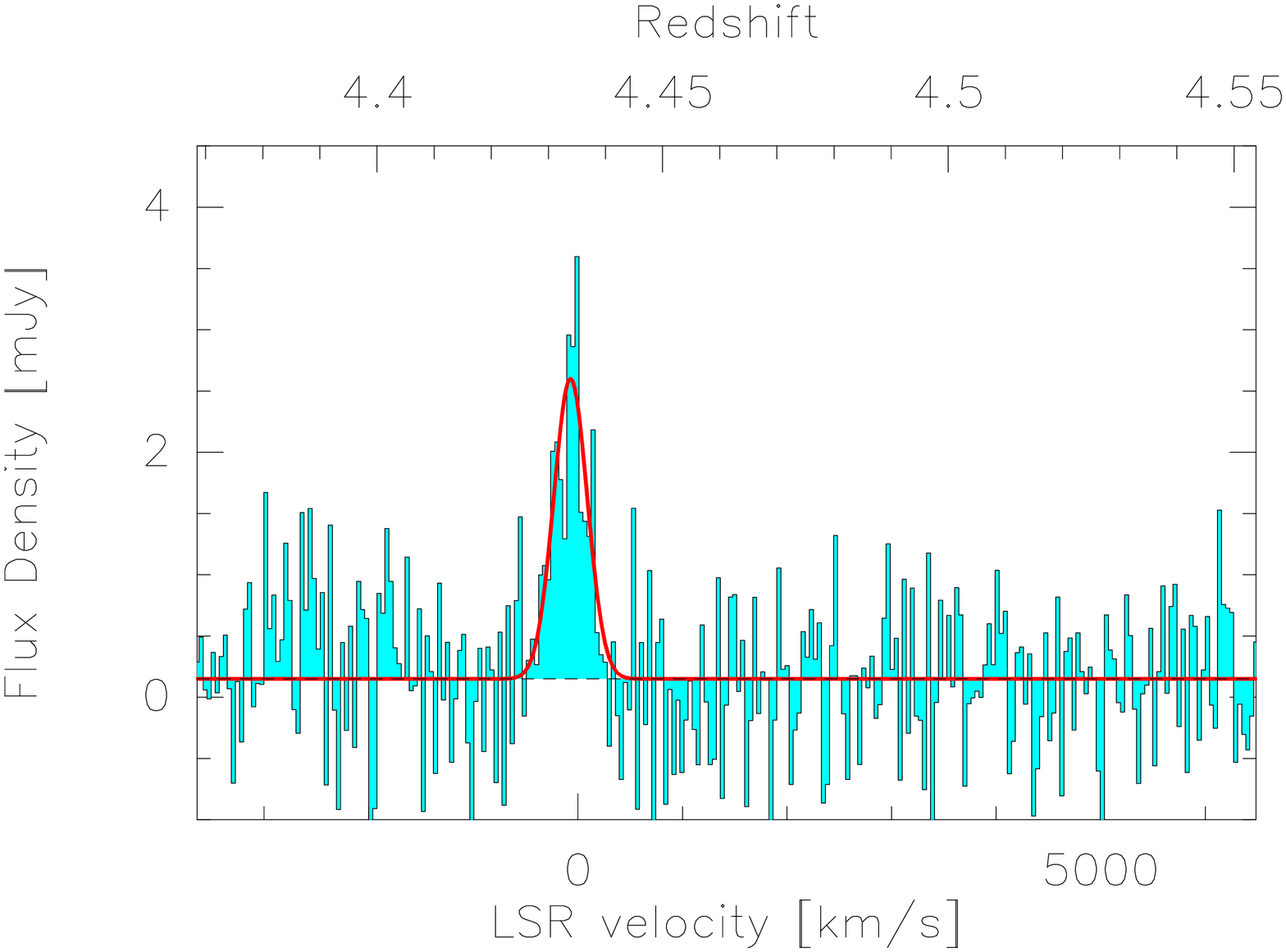} 
\includegraphics[scale=0.35]{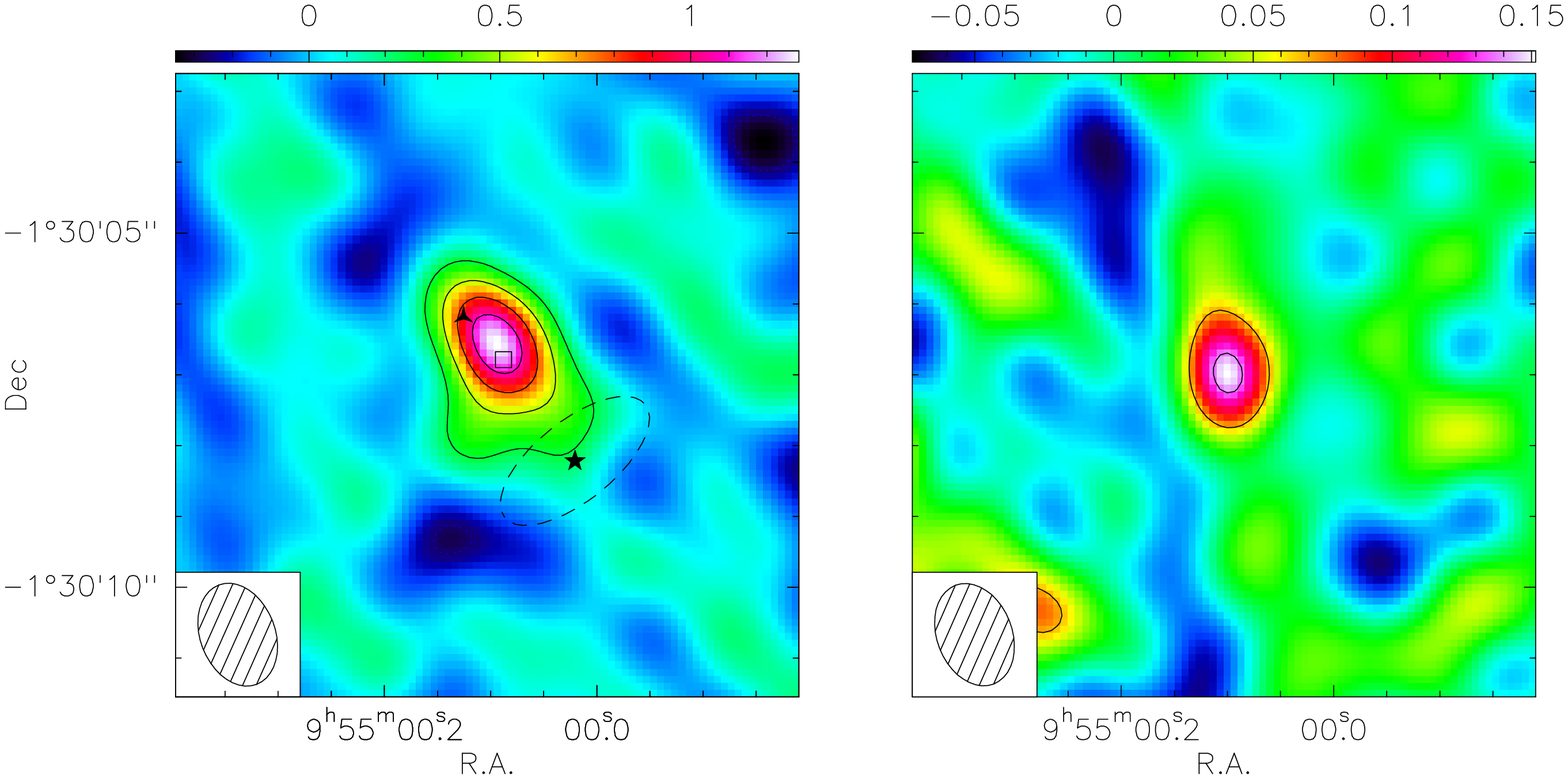}
\caption{Left panel: CO(5--4) spectrum of B0952 obtained with the PdBI by integrating over a region which follows the 2.5$\sigma$ contour on the core map. The spectrum is shown on top of a continuum of $F_{cont}=0.15\pm 0.03$~mJy (see right panel). Middle panel: map of the CO(5--4) line emission region integrated over the velocity range $\rm -396 < v < 261~[km~s^{-1}]$. Contour levels are shown in steps of 2.5$\sigma$, where 1$\sigma=0.07$ Jy~km~s$^{-1}$~beam$^{-1}$. The synthesized beam of $1.50''\times 1.06''$ is shown in the bottom-left insets. Right panel: map of the continuum emission obtained from the line-free channels of the 3.58~GHz wide spectrum. Contour levels are shown in steps of 2.5$\sigma$, where 1$\sigma=0.03$ mJy~beam$^{-1}$.} 
\label{figCO}
\end{figure}

\begin{figure}
\centering
\includegraphics[scale=0.25]{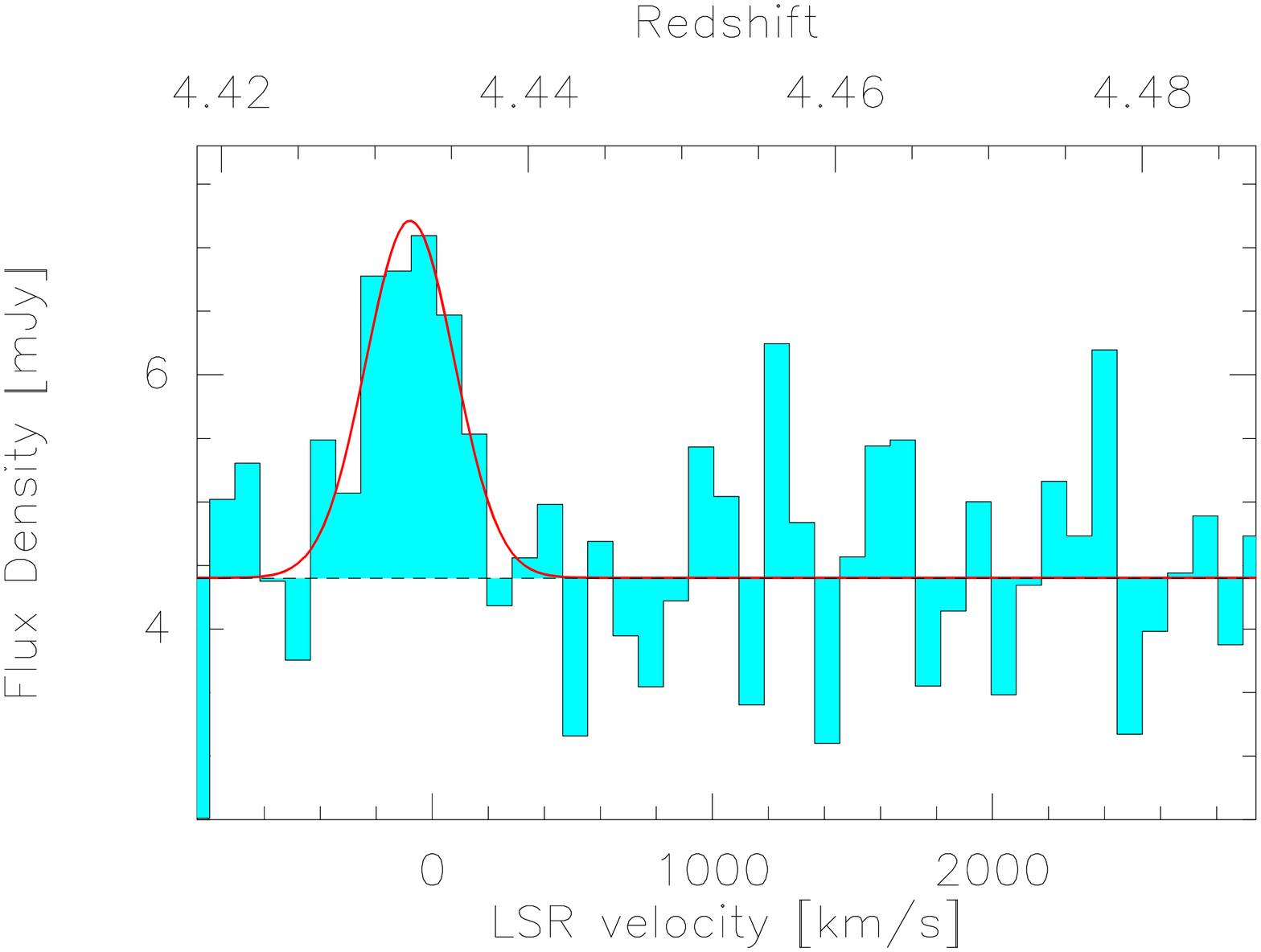} 
\includegraphics[scale=0.35]{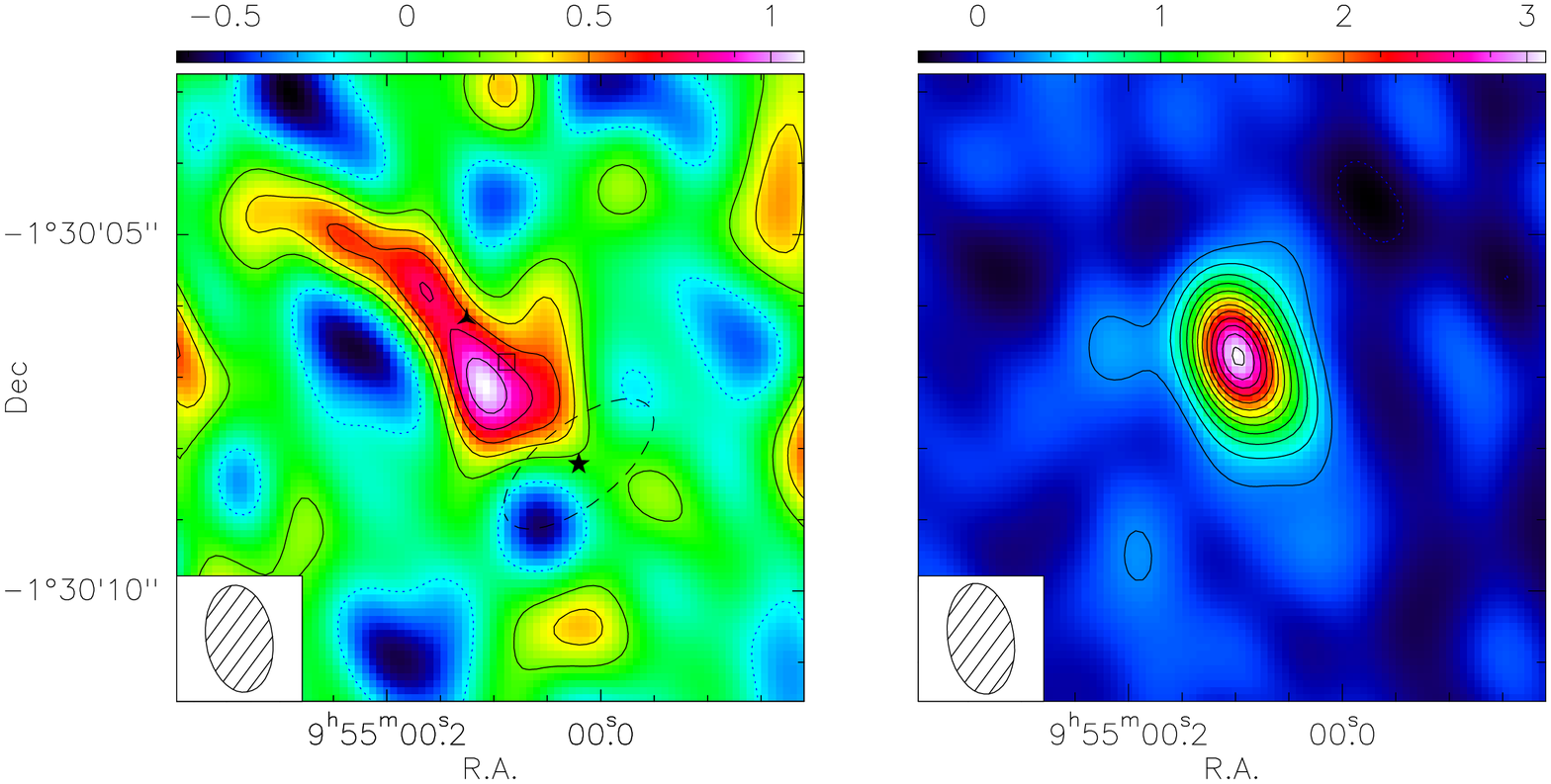}
\caption{Left panel: [NII] spectrum of B0952 obtained with the PdBI by integrating over a region which follows the 3$\sigma$ contour on the core map. The spectrum is shown on top of a continuum of $F_{cont}=4.4\pm 0.1$~mJy (see right panel). Middle panel: map of the [NII] line emission region integrated over the velocity range $\rm -390 < v < 150~[km~s^{-1}]$. Contour levels are shown in steps of 1$\sigma=1$ Jy~km~s$^{-1}$~beam$^{-1}$. The synthesized beam of $1.53''\times 0.96''$ is shown in the bottom-left insets. Right panel: map of the continuum emission obtained from the line-free channels of the 3.4~GHz wide spectrum. Contour levels are shown in steps of 3$\sigma$, where 1$\sigma=0.08$ mJy~beam$^{-1}$.} 
\label{figNII}
\end{figure}

The CO(1-0) emission line luminosity $L_{\rm CO(1-0)}$ provides constraints on the molecular hydrogen mass of galaxies. From our observations of the CO(5-4) emission line (Fig. \ref{figCO}) we obtain a de-lensed $L_{\rm CO(5-4)}=$, which converts into $M_{\rm H_2}=2\times 10^9 M_{\odot}$, assuming a constant brightness in the CO(5-4) and CO(1-0) transitions. Moreover, the width of the CO(5-4) line and the upper limit on the size of the [CII] emission region (G12) allows us to constrain the B0952 dynamical mass $M_{dyn}<2.4\times 10^{10} M_{\odot}$, and to infer its stellar mass $M_*=M_{\rm dyn}-M_{\rm H_2}<2.2\times 10^{10} M_{\odot}$. Since B0952 is powered by a super-massive black hole having $M_{\rm BH}=2\times 10^9 M_{\odot}$ (Shields et al. 2006), the $M_*$ value that we found is significantly smaller than what we would infer from the local $M_{\rm BH}-M_*$ relation, i.e. $M_*\sim 10^{12} M_{\odot}$. We note that similar deviations from the local $M_{\rm BH}-M_*$ relation have been found also in other high-z sources (Wang et al. 2010), suggesting that black holes accretion may be more efficient at early epochs.\\
%(luminosity and width) are consistent with a molecular hydrogen mass $M_{\rm H_2}\sim 2\times 10^9 M_{\odot}$, and a dynamical mass $M_{\rm dyn}< 2.4\times 10^{10} M_{\odot}$. We therefore 
The measurement of the [NII]/[CII] flux ratio is a powerful method to determine the gas metallicity in galaxies at all redshifts (Nagao et al. 2011). We combine our detection of the [NII] line in B0952 (Fig. \ref{figNII}) with our results on its [CII] line to measure a flux ratio [NII]/[CII]=0.055$\pm$0.015. By exploiting the calculations by Nagao et al. (2012), we infer a gas metallicity $\rm log~\rm (Z_{\rm gas}/Z_{\odot})=-0.04\pm 0.44$. In particular, the [NII] map (Fig. \ref{figNII}, middle panel) shows a plume of emission extending towards the North-West part of the map which is characterized by a metallicity $Z>Z_{\odot}$. The high metallicity that we infer in this region can not be due to {\it in-situ} star formation, since it has not been detected in the optical, but it is more likely the result of metals ejected from the QSO host galaxy through SN or QSO-driven winds.
\section{[CII] emission in S1148}
\begin{figure}
\centering
\includegraphics[scale=0.25]{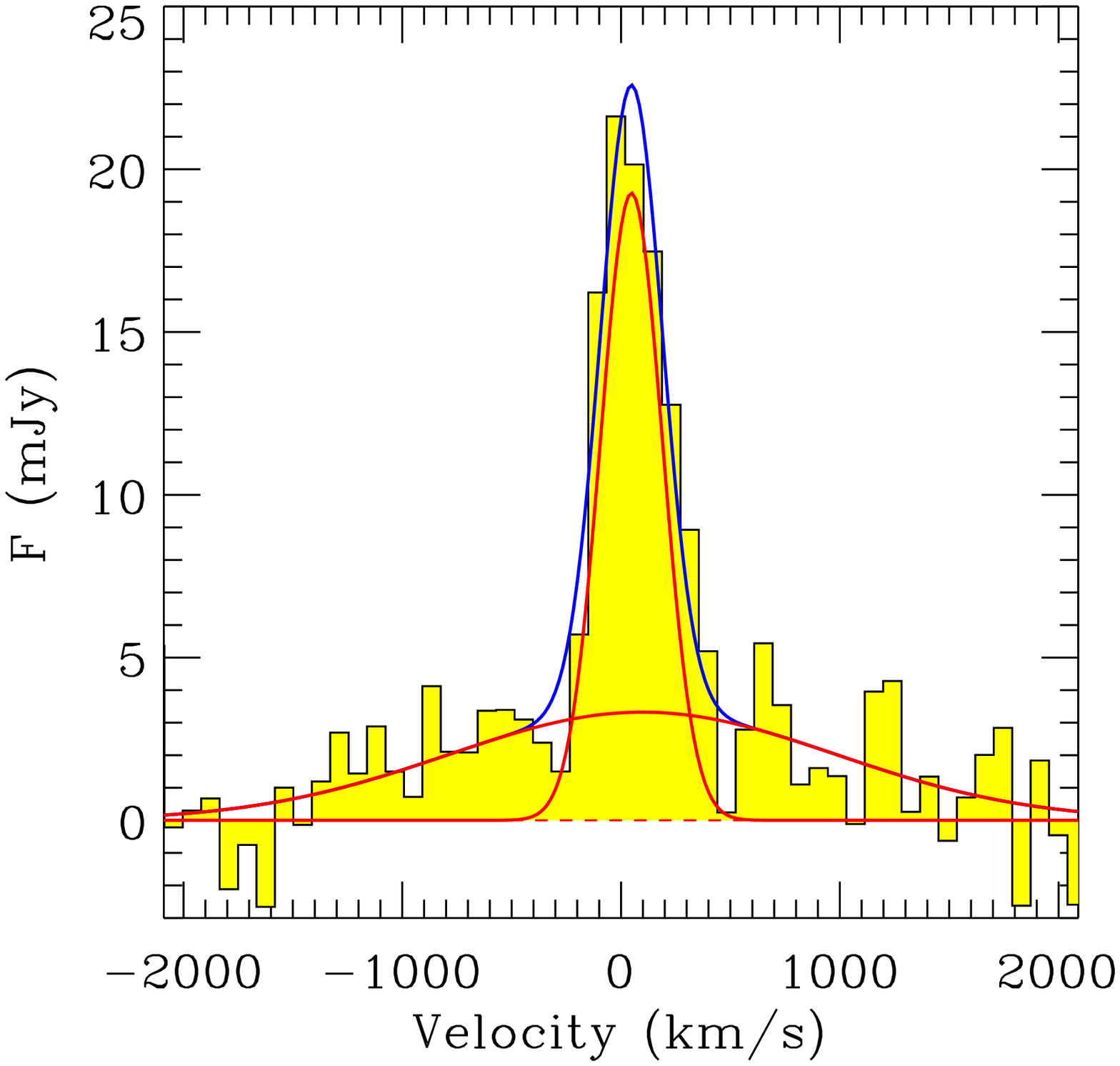} 
\includegraphics[scale=0.25]{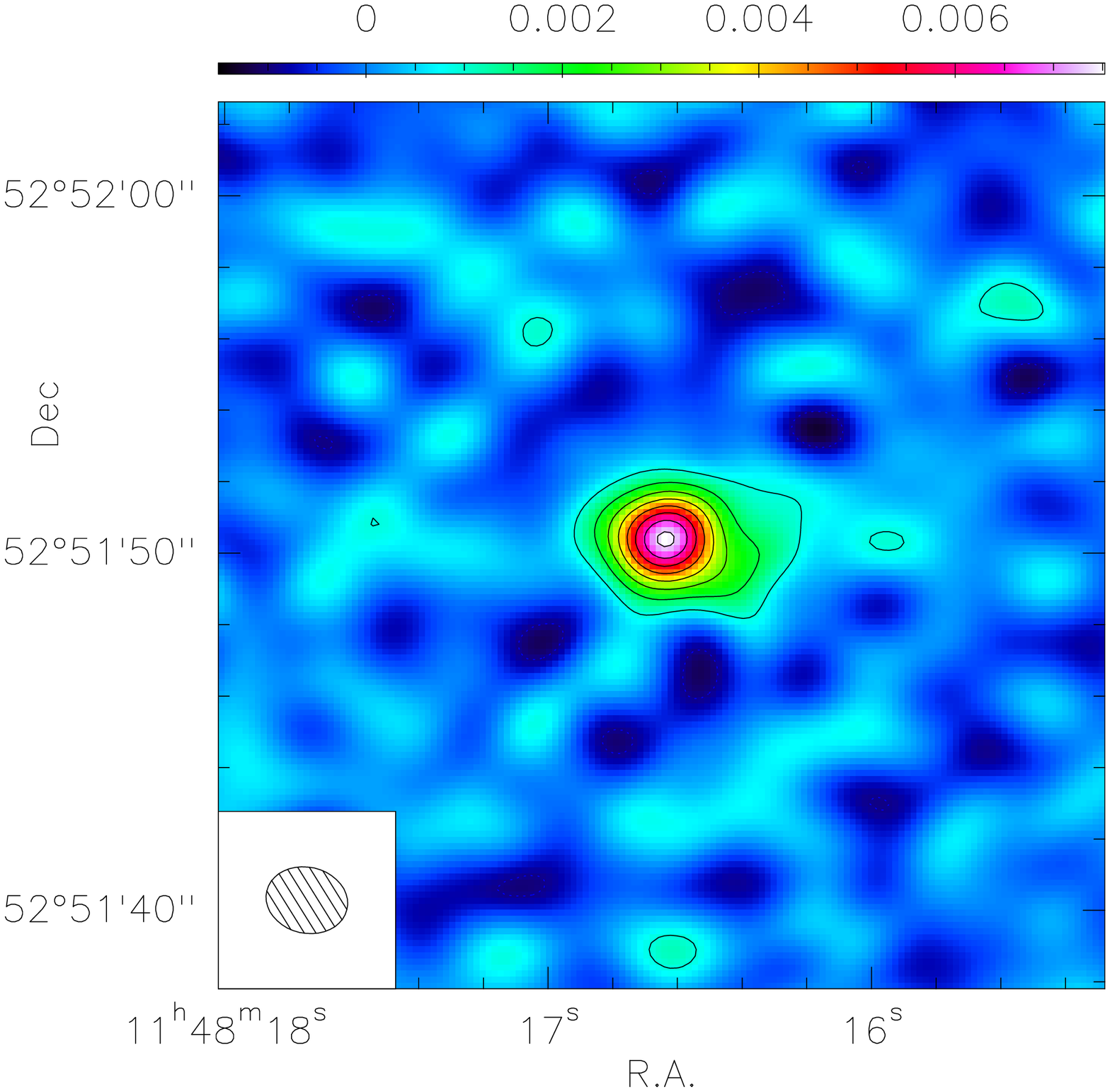}
\includegraphics[scale=0.25]{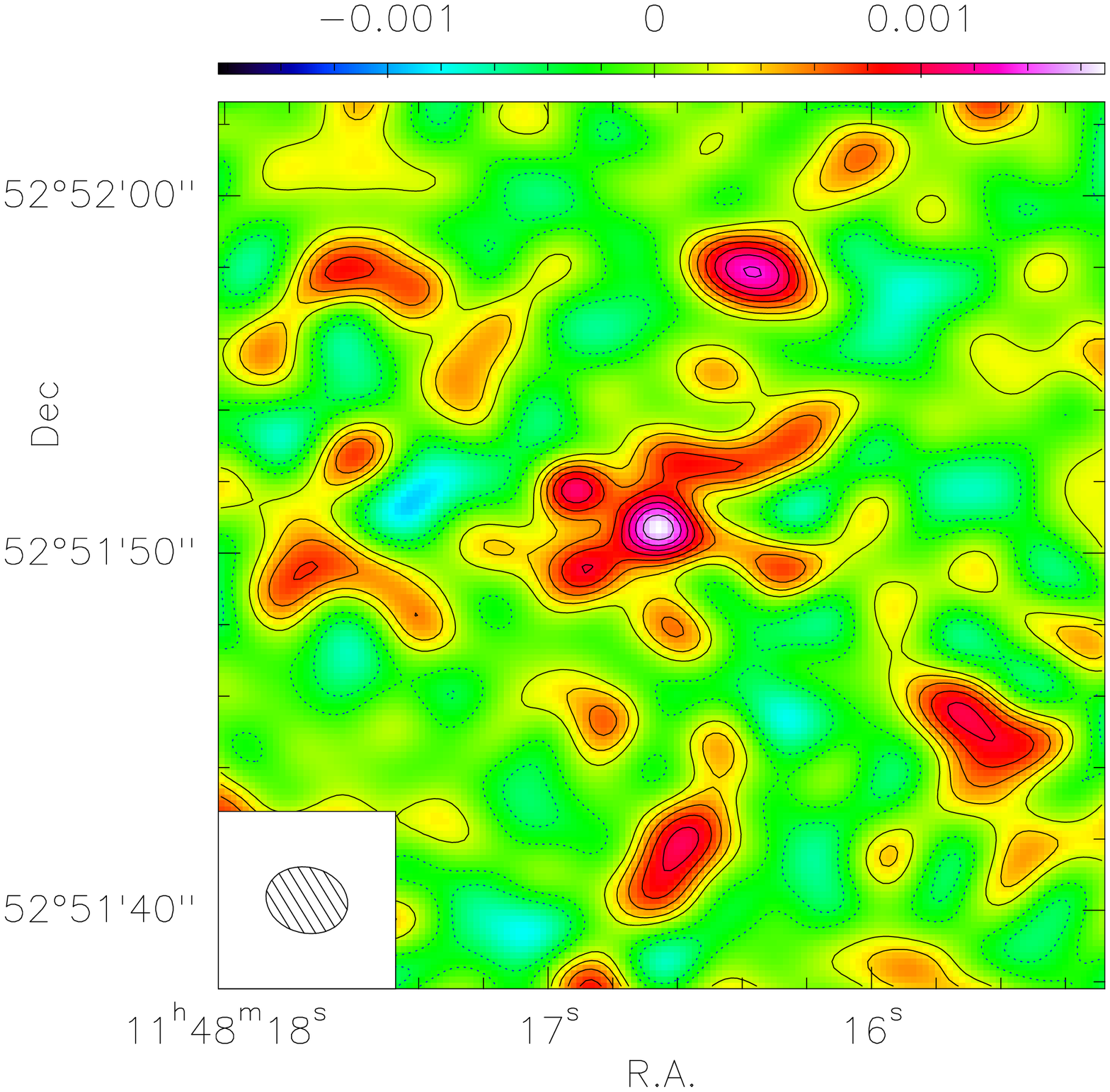}
\caption{Left panel: [CII] spectrum of S1148 obtained with the PdBI extracted from an aperture with a diameter of 6$''$. The red lines show a double Gaussian fit (FWHM=345~km~s$^{-1}$ and FWHM=2030~km~s$^{-1}$) to the line profile, while the blue line shows the sum of the two Gaussian components. Middle panel: map of the [CII] line narrow component ($\rm -300<v<+400~km~s^{-1}$). Levels are in steps of  $\rm 0.64~Jy~km~s^{-1}~beam^{-1}$ (i.e. 3$\sigma$). The beam of the observation is shown in the bottom-left corner. Right panel: map of the [CII] line wings ($\rm -1300<v<-300~km~s^{-1}$ and $\rm +500<v<+1300~km~s^{-1}$). Levels are in steps of $1\sigma=\rm 0.36~Jy~km~s^{-1}~beam^{-1}$.} 
\label{figwings}
\end{figure}

We detect broad wings in the [CII] emission line in S1148 at z=6.4 (Fig. \ref{figwings}). The extent of the wings ($\sim$1300~$\rm km~s^{-1}$) and the size of the [CII] emitting region associated to the wings ($\sim 16$~kpc), are indicative of a QSO-driven massive outflow with the highest outflow rate ever found ($\rm \dot{M}> 3500~M_{\odot}~yr^{-1}$). This lower limit on the outflow rate is higher than the SFR in the QSO host ($\rm SFR\approx 3000~M_{\odot}~yr^{-1}$, Bertoldi et al. 2003), thus implying that the gas content in the host galaxy will be cleaned, and therefore star formation will be quenched, in less than $\rm 6\times 10^6~yrs$. Such a fast and efficient quenching mechanism, already at work at z$>$6, is what is required by models to explain the properties of massive, old and passive galaxies observed in the local Universe and at $\rm z\sim 2$ (Cimatti et al. 2004).
\section{Conclusions}
We have presented PdBI observations of B0952 (a lensed QSO at z=4.4) and S1148 (one of the most distant QSO known, at z=6.4). For what concerns B0952, we have obtained one of the first resolved map of [CII] at high-z. Our data allow us to reveal the presence of a component which eludes optical observations, and which is likely a companion galaxy in the phase of merging with the QSO host. From the observed CO(5-4) emission line we find that B0952 deviates from the local $M_{\rm BH}-M_*$ relation, being characterized by a smaller stellar mass with respect to local galaxies hosting black holes with masses similar to the B0952 one. Our result confirms the conclusions of previous studies, according to which black holes accretion may be more efficient at high-z than in the local Universe. We have also detected the [NII] emission line in B0952, whose flux density is consistent with a solar metallicity of the gas. The morphology of the [NII] emission is also suggestive of a SN/QSO-driven outflow. In S1148, we have detected broad wings in the [CII] emission line, indicative of gas which is outflowing from the host galaxy. In particular, properties of the line and the extension of the [CII] emission coming from the wings, are indicative of a QSO-driven massive outflow (Valiante et al. 2012) with the highest outflow rate ever found. These observations underline the importance of millimeter observations of high-z star-forming galaxies for constraining the properties of their ISM, and for studying processes related to galaxy formation, as galaxy merging and SN/QSO feedback.\\
\label{conclus}

\end{document}